\documentclass[12pt]{article}

\usepackage{amssymb}
\usepackage{epsfig}
\usepackage{amsmath}

\begin{document}

\title{Signs of universality in the behavior of elastic $\textit{pp}$ scattering cross-sections at high energies}

\author{A. P. Samokhin \\
\textit{A.A. Logunov Institute for High Energy Physics}\\
\textit{of NRC ``Kurchatov Institute''}\\
\textit{Protvino, 142281, Russian Federation}}

\date{}

\maketitle

\begin{abstract}
We give a phenomenological analysis of the behavior of  inelastic,  elastic and total cross-sections  of the
high energy $\textit{pp}$ interaction. In particular, we argue that the universal picture of  behavior of   cross-sections and their ratios is a consequence of the rapid increase of  inelastic  cross-section with energy and its large value compared to $\sigma_{\mathrm{el} }(s)$. We observed that the value of  the fundamental ratio $(m_{\pi^{\mathrm{0}}}/m_{p}) $, the minimum value of the ratio $ (\sigma_{\mathrm{el}}/\sigma_{\mathrm{tot}})$, and some other quantities are determined by the roots of the  equation $ (9\,x^{2}+4\,\sqrt{2}\,x-1)=0 $.
\end{abstract}


\section{Introduction}
Elastic hadron-hadron scattering at high energies is a unique phenomenon of its kind, which has no analogue in neither classical nor quantum mechanics.
Unlike inelastic processes,
elastic scattering has no probabilistic interpretation; we can only talk about
the probability of survival of colliding hadrons (elastic scattering plus passage
without interaction). This fact does not allow us to trust simple visual models of elastic hadron-hadron scattering.
Elastic scattering is closely related to inelastic processes in this channel. In particular, if we assume that all inelastic processes are absent, then the amplitude of elastic scattering will be zero [1]. Moreover, due to the unitarity condition, the properties of the elastic scattering amplitude are largely
determined by the totality of  the inelastic processes in this channel. In this sense, elastic scattering is a shadow of the particle production processes [2], [3].

For these reasons, elastic hadron-hadron  scattering is one of the most difficult
problems in high energy physics, and we do not yet have an adequate model
for this phenomenon. Indeed, all important experimental discoveries in this
field, such as the growth of $\sigma_{\mathrm{el}}(s)$, $\sigma_{\mathrm{tot}}(s)$, $\sigma_{\mathrm{el}}(s)/\sigma_{\mathrm{tot}}(s) $, the existence of the second
diffraction cone, were unexpected, surprising for us.

High-energy inelastic  processes exhibit some features of collective behavior and universality due to the evolution of quark-gluon matter formed during \textit{pp}  collisions into hadrons (see
e.g. [4], [5] for a recent review). As a shadow process, elastic scattering should also have some properties of universality. Indeed, there is geometric scaling at the ISR energies [6--8], stationary points of the differential cross-section in the ISR-LHC energy range [9--10], and energy independence of the ratio of bump to dip positions at the ISR-LHC energies [11--12].
In this note we discuss some other features of universality in the behavior of the   \textit{pp} cross-sections.

We argue that the increase of
$\sigma_{\mathrm{el}}(s)$, $\sigma_{\mathrm{tot}}(s)$, $\sigma_{\mathrm{el}}(s)/\sigma_{\mathrm{tot}}(s) $ at high energies is caused by the monotonic and fast (faster than $ \ln (\sqrt{s} \,)$) growth of the inelastic cross-section $\sigma_{\mathrm{inel} }(s)$ and its large value  compared to $\sigma_{\mathrm{el} }(s)$. This fact leads to a monotonous increase in the difference $(\sigma_{\mathrm{inel}}(s)-\sigma_{\mathrm{el}}(s)) $ and to the nontrivial behavior of elastic and total  cross-sections, their ratio and the intensity of inelastic interaction, ($\sigma_{\mathrm{el}}/\sigma_{\mathrm{tot}}$)(1 - $\sigma_{\mathrm{el}}/\sigma_{\mathrm{tot}}$), with increasing energy. All these quantities have minima and reach them at certain energy values that satisfy the inequalities $\sqrt{s_{\mathrm{tot}}} < \sqrt{s_{\mathrm{el}}} < \sqrt{s_{\mathrm{rat}}}$. The ratio $\sigma_{\mathrm{el}}/\sigma_{\mathrm{tot}}$ and the intensity of inelastic interaction
reach their minima at the same energy value $\sqrt{s}=\sqrt{s_{\mathrm{rat}}}$. This picture of behavior of cross-sections and their ratios is universal, i.e. it is valid for any hadron-hadron collision, and confirms the shadow nature of elastic scattering. The minimum value of
the ratio $\sigma_{\mathrm{el}}/\sigma_{\mathrm{tot}}$ for \textit{pp} collisions is reached at $\sqrt{s}\simeq$ 30 GeV and is equal to $(0.1747 \pm 0.0012)$ [24]. We observed that if the minimum value of the intensity of inelastic interaction is $ \xi_{\mathrm{1}}\equiv(m_{\pi^{\mathrm{0}}}/m_{p})\approx0.143856$, then
the predicted minimum value of
the ratio $\sigma_{\mathrm{el}}/\sigma_{\mathrm{tot}}$ will be 0.1742, i.e. very close to the experimental value. Current experimental data also do not exclude that $\xi_{1}$ may be an upper bound for the ratio of the real to imaginary part of the
elastic scattering amplitude in the forward direction, $ \rho(s)$. We see that the values of many dimensionless observables in \textit{pp} collisions are not accidental, but are related to the  parameter $\xi_{1}$. In this sense, the role of the parameter $\xi_{1} $ in hadron physics is similar to the role of the Golden Ratio (GR) in macroscopic
phenomena. Therefore, it is quite possible that $\xi_{1}$ is, like GR, the root of a quadratic equation. Indeed, we found that one of the roots of the quadratic equation $(9\,x^{2}+4\,\sqrt{2}\,x-1)=0$ is very close to $\xi_{1}$, (the accuracy is $(\xi_{1}-x_{1})\approx 10^{-6}$). Surprisingly, the second root of this equation, $-\xi_{2}\approx -0.77239 $, is also related to the hadron mass ratios ($\rho$ meson, proton, $ \Delta$ baryon) and    gives the   effective threshold values for
the elastic \textit{pp} scattering amplitude. We do not know why the roots of this equation, the scales $ \xi_{1}, \xi_{2}$, are relevant for hadron physics, but we believe that this observation
may help to find an adequate model for the elastic scattering amplitude.

In Section 2, we will discuss the  properties of cross-sections and the  role of dimensionless parameters.  A brief summary and discussion are given in Section 3.

\section{Cross-sections and dimensionless parameters}

 The energy dependence of  elastic $\sigma_{\mathrm{el}}(s)$, inelastic $\sigma_{\mathrm{inel}}(s)$ and total cross-sections $ \sigma_{\mathrm{tot}}(s)=\sigma_{\mathrm{el}}(s)+\sigma_{\mathrm{inel}}(s)$ for \textit{pp}  collisions  is shown in Fig. 1. The inelastic  cross-section grows monotonically with energy $ \sqrt{s} $. This growth
of $\sigma_{\mathrm{inel}}(s)= \sum\limits_{n,\,\mathrm{inel}}^{N(s)} \sigma_{n}(s)$ can be formally attributed to a huge number of open inelastic 
channels $N(s)$ [13]. As can be seen from Fig. 1, the inelastic cross-section
grows faster than $ \ln (\sqrt{s})$. One of the first models of hadronization of the  collision energy [14] gives  $ \sigma_{\mathrm{inel}}(s)\simeq{\mathrm{ln}}^{2}(\sqrt{s}\,)$ and according to  Froissart-Martin's bound [15], [16] we have $ \sigma_{\mathrm{inel}}(s)\leq  C {\mathrm{ln}}^{2}(\sqrt{s}\,)$. At low energies, the  elastic cross-section $\sigma_{\mathrm{el}}(s)=\int dt(d\sigma/dt) $  decreases with energy (scattering due to the short-range potential of the strong hadron-hadron interaction). For a certain value of $ s = s_{\mathrm{tot}}, $ the  positive value of $\sigma_{\mathrm{inel}}^{'}$ compensates for  the negative value of $\sigma_{\mathrm{el}}^{'}$ (prime denotes the derivative with respect to $ {\mathrm{ln}}(\sqrt{s}\,)$). That is, $\sigma_{\mathrm{tot}}^{'}=(\sigma_{\mathrm{el}}^{'}+\sigma_{\mathrm{inel}}^{'})=0$ at $s=s_{\mathrm{tot}}$, and the total cross-section  reaches its minimum at $s=s_{\mathrm{tot}}$.
From Fig. 1 we see that for \textit{pp}  collisions $ (\sqrt{s_{\mathrm{tot}}})_{pp} \simeq $ 10 GeV.
\begin{figure}[t]
\centering
\includegraphics[height=8.5cm]{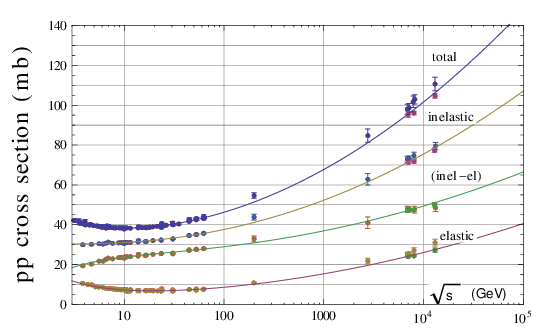}
\caption{ Cross-sections for   \textit{pp}
collisions as a function of total center-of-mass energy  $\sqrt{s}$.  The experimental
data for $\sigma_{\mathrm{tot}}(s)$ and $\sigma_{\mathrm{el}}(s)$ are from [21-23], lines represent a fit according to the form: $ \sigma_{\mathrm{el,tot}}=C_{2} x^{2}+C_{1} x+C_{0}+C_{4} e^{-x} , \, x=\ln(\sqrt{s})$. Inelastic cross-section and $ \Delta$ are given by    $\sigma_{\mathrm{inel}}=(\sigma_{\mathrm{tot}}-\sigma_{\mathrm{el}})$
and $\Delta=(\sigma_{\mathrm{inel}}-\sigma_{\mathrm{el}})=(\sigma_{\mathrm{tot}}-2\,\sigma_{\mathrm{el}})$ both for data and for fits.}
\end{figure}
For $s>s_{\mathrm{tot}}$, the total cross-section increases with energy, but the elastic one continues to decrease.
Due to the shadow nature of elastic scattering, the rapid growth of the inelastic cross-section $\sigma_{\mathrm{inel}}(s)$ with energy causes the elastic cross-section $\sigma_{\mathrm{el}}(s)$ to grow. Indeed, if we assume that $ \sigma_{\mathrm{inel}}(s)\simeq{\mathrm{ln}}^{1+\epsilon}(\sqrt{s}\,)$ then according to the well-known bound [17],[18] (which follows from the general principles of QFT with a mass gap)
\begin{equation}
\sigma_{\mathrm{tot}}^{2}(s) \leq \sigma_{\mathrm{el}}(s) (\frac{4\pi}{t_{0}}){\mathrm{ln}}^{2}(\frac{s}{s_{0}}) ,\,\,\,\,\,\,\,  t_{0}\sim 4m_{\pi}^{2} \,,
\end{equation}
the elastic cross-section will satisfy  $ \sigma_{\mathrm{el}}(s)\geq C {\mathrm{ln}}^{2\epsilon}(\sqrt{s}\,)$, that is, it should grow at high energies [19]. Therefore, for a certain value of $ s = s_{\mathrm{el}} > s_{\mathrm{tot}} $, the elastic cross-section $\sigma_{\mathrm{el}} $ should reach its minimum value (for \textit{pp} collisions  $ (\sqrt{s_{\mathrm{el}}})_{pp} \simeq $  20 GeV, see Fig. 1), and then it should increase with energy.

At all energies  of $ \sqrt{s}\geq $ 3 GeV, the inelastic cross-section is much larger than the elastic one. Moreover, as can be seen from Fig. 1, the difference
between inelastic and elastic cross-sections
\begin{equation}
\Delta(s)=\sigma_{\mathrm{inel}}(s)-\sigma_{\mathrm{el}}(s)
\end{equation}
increases monotonically with energy (see Fig. 1), that is, $ \Delta^{'}>0.$ This means that the  inelastic cross-section grows faster than the  elastic one, $\sigma_{\mathrm{inel}}^{'} > \sigma_{\mathrm{el}}^{'}$. We see also (see Fig. 1) that at all energies $\Delta(s)>  \sigma_{\mathrm{el}}(s) $ [20].

The ratio $ \sigma_{\mathrm{el}}(s)/\Delta(s) $ decreases with energy up to $ \sqrt{s}=  \sqrt{s_{\mathrm{el}}} $ due to the decrease of $ \sigma_{\mathrm{el}}(s) $ and the growth of $ \Delta(s)$. However, at higher energies, this ratio grows because $\Delta > \sigma_{\mathrm{el}},\, \sigma_{\mathrm{el}}^{'} \gtrsim \Delta^{'} $ and therefore $ (\sigma_{\mathrm{el}}/\Delta)^{'} > 0$. So, as can be seen from
\begin{equation}
\frac{\sigma_{\mathrm{tot}}}{\sigma_{\mathrm{el}}}=1+\frac{\sigma_{\mathrm{inel}}}{\sigma_{\mathrm{el}}}=2+\frac{\Delta}{\sigma_{\mathrm{el}}} \, ,
\end{equation}
the ratios $ \sigma_{\mathrm{el}}/\Delta,  \,\sigma_{\mathrm{el}}/\sigma_{\mathrm{inel}} $ and $  \sigma_{\mathrm{el}}/\sigma_{\mathrm{tot}} $ reach their minima at the same value of $ s=s_{\mathrm{rat}} > s_{\mathrm{el}}$. From Fig. 2 we see that for \textit{pp}  collisions $ (\sqrt{s_{\mathrm{rat}}})_{pp} \simeq $ 30 GeV.
\begin{figure}[t]
\centering
\includegraphics[height=8.5cm]{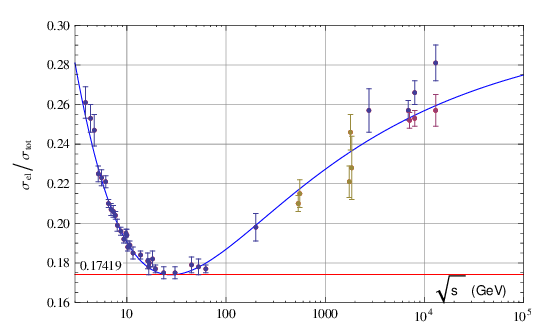}
\caption{The elastic to total cross-section ratio for \textit{pp} collisions as a function of energy $\sqrt{s}$. The experimental
data for $\sigma_{\mathrm{el}}$ and $\sigma_{\mathrm{tot}}$ are from Refs. [21--23],
including the \textit{$\bar{p}$p} data at $\sqrt{s}=546$ and 1800 GeV [21].
The line represents a ratio of  the fits for $\sigma_{\mathrm{el}}$ and $\sigma_{\mathrm{tot}}$ (see Fig. 1 ).}
\end{figure}
\begin{figure}[t]
\centering
\includegraphics[height=8.5cm]{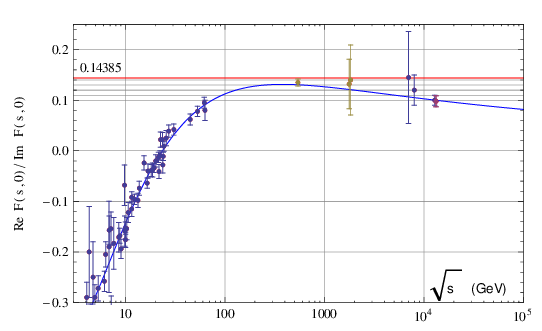}
\caption{ Ratio of the
real to imaginary part of the elastic \textit{pp} scattering amplitude in the forward direction as a function of energy $\sqrt{s}$ (the experimental
data are from Refs. [21--22], including the \textit{$\bar{p}$p} data at $\sqrt{s}=546$ and 1800 GeV [21]).
The line represents a fit according to the form: $ \rho=(a_{0} +a_{1}x+(a_{2} +a_{3}x) e^{-x})/ \sigma_{\mathrm{tot}},\, x=\ln(\sqrt{s}) $, where the fit from Fig. 1 was used for $\sigma_{\mathrm{tot}}$. }
\end{figure}
Note that all the properties of cross-sections and their ratios discussed here, including inequalities
\begin{equation} s_{\mathrm{tot}} < s_{\mathrm{el}} < s_{\mathrm{rat}} \, ,
 \end{equation}
have a universal character, that is, they are valid not only for \textit{pp}, but also for any hadron-hadron collision (see plots for \textit{$\bar{p}$p} , $ \pi p $ and $ K p$ in [21]). All these results confirm the shadow nature of elastic scattering.

The rapid growth of $ \sigma_{\mathrm{inel}}(s)$ and growth of the ratio $ \sigma_{\mathrm{el}}/\sigma_{\mathrm{tot}}$ at high energy are a consequence of the increase of the inelastic interaction intensity  [20]
\begin{equation}
\frac{\sigma_{\mathrm{inel}}(s)} {16\pi B(s)} \simeq \frac{\sigma_{\mathrm{el}}(s) \sigma_{\mathrm{inel}}(s)}{\sigma_{\mathrm{tot}}^{2}(s)}\,,
 \end{equation}
where $B(s)$ is the slope of the forward diffraction peak. The increase of the ratio (5) means that $ \sigma_{\mathrm{inel}}(s)$ grows faster than the cross-sectional area of the interaction region, which is proportional to $B(s)$. It is obvious  (since $\sigma_{\mathrm{inel}}=\sigma_{\mathrm{tot}}-\sigma_{\mathrm{el}}$) that the intensity of inelastic interaction (5) grows when the ratio $ \sigma_{\mathrm{el}}/\sigma_{\mathrm{tot}}$ grows and vice versa; like $ \sigma_{\mathrm{el}}/\sigma_{\mathrm{tot}}$,  it reaches its minimum value at $s=s_{\mathrm{rat}}$.   It is also obvious that the upper bound of the intensity of inelastic interaction (5) is 0.25. Let's assume that the lower bound is $ \xi_{1}$, i.e.
\begin{equation}
\xi_{\mathrm{1}}\leq(\frac{\sigma_{\mathrm{el}}\sigma_{\mathrm{inel}}}{\sigma_{\mathrm{tot}}^{2}})\leq \frac{1}{4},\,\, \mathrm{where} \,\,\,   \xi_{\mathrm{1}}\equiv \frac{m_{\pi^{\mathrm{0}}}} {m_{p}} \approx0.143856.
\end{equation}
Then the predicted minimum value
of the ratio $ \sigma_{\mathrm{el}}/\sigma_{\mathrm{tot}}$ (and the maximum for $ \sigma_{\mathrm{inel}}/\sigma_{\mathrm{tot}}$) will be equal to
\begin{equation}
\frac{\sigma_{\mathrm{el}}}{\sigma_{\mathrm{tot}}}\geq(\frac{1}{2}-(\frac{1}{4}-\xi_{\mathrm{1}})^{0.5})\approx0.1742 \, ,
\end{equation}
\begin{equation}
\frac{\sigma_{\mathrm{inel}}}{\sigma_{\mathrm{tot}}}\leq(\frac{1}{2}+(\frac{1}{4}-\xi_{\mathrm{1}})^{0.5})\approx0.8258 \, .
\end{equation}
This is very close to the experimental value for \textit{pp} collision [24] (see Fig. 2)
\begin{equation}
(\sigma_{\mathrm{el}}/\sigma_{\mathrm{tot}})_{\mathrm{min}}^{\mathrm{exp}}=0.1747\pm0.0012 \, .
\end{equation}
From (7) it is easy to find the minimum value for other ratios, for example,
\begin{equation}
\sigma_{\mathrm{el}}\geq0.2109\,\sigma_{\mathrm{inel}} \, .
\end{equation}

We see that the dimensionless
parameter $ \xi_{1}$ determines not only the ratio of the masses of the lightest meson and the lightest baryon but also the minimum values of the intensity of inelastic interaction and  $(\sigma_{\mathrm{el}}/\sigma_{\mathrm{tot}})_{pp}$. Current experimental data also do not exclude that $\xi_{1}$ may be an upper bound for the ratio of the real to imaginary part of the
elastic scattering amplitude in the forward direction, $ \rho(s)$, (see Fig. 3)
\begin{equation}\rho_{pp}(s)=\frac{\mathrm{Re}\,F(s,t=0)}{\mathrm{Im}\,F(s,t=0)} \leq \xi_{1} \, .
\end{equation}

Let's note that the minimum values of $ (\sigma_{\mathrm{el}}/\sigma_{\mathrm{tot}})$ and of the intensity of inelastic interaction for meson-proton collisions are less than for \textit{pp} one [25]. For example, $ (\sigma_{\mathrm{el}}/\sigma_{\mathrm{tot}})_{\mathrm{min}}^{\pi p} \simeq$ 0.14. This corresponds to the value $(5\, \xi_{1}/6)$ for the intensity of inelastic interaction in $\pi p$ collisions,  however, the errors in experimental data [21] are too large for an unambiguous conclusion.

The role of the parameter $\xi_{1} $ in hadron physics is similar to the role of the GR (the roots of the equation $x^{2}+x-1=0 $) in macroscopic
phenomena. Therefore, it is not so surprising that $\xi_{1} $ is one of the roots of the quadratic equation
\begin{equation}
9\,x^{2}+4\,\sqrt{2}\,x-1=0.
\end{equation}
Unlike GR, where there is essentially one scale (since the product of the roots of the equation is -1), the roots of this equation give us two different parameters
\begin{equation}
x_{\mathrm{1}}=\frac{\sqrt{17}-2\,\sqrt{2}}{9}\simeq0.143853\simeq\xi_{\mathrm{1}}\equiv(m_{\pi^{\mathrm{0}}}/m_{p})\approx0.143856 \, ,
\end{equation}
\begin{equation}
x_{\mathrm{2}}=\frac{\sqrt{17}+2\,\sqrt{2}}{(-9)}\equiv-\xi_{\mathrm{2}}\simeq-0.77239.
\end{equation}
We see that $(\xi_{1}-x_{1})\approx 10^{-6}$,  so we can say that $\xi_{1}$ is the root of  equation (12).
The parameters $\xi_{1}, \,\xi_{2}$ have the following properties: $ 9 \, \xi_{1} \xi_{2}$ = 1, $(\xi_{2}-\xi_{1})\approx 0.62854$, (GR $\approx$ 0.61803), $(\xi_{1}+\xi_{2})\approx 0.91625$, (Catalan's constant $G \approx 0.91597$). Interestingly, $ (0.5 (G-\mathrm{GR})-(m_{\pi^{\mathrm{\pm}}}/ m_{p}))\approx 10^{-4}$.

The value of the second parameter, $  \xi_{2}$, is meaningful for hadron physics, since $\xi_{2} \, m_{p}\approx 0.725 $ GeV is close to the mass of the lightest vector meson, and $m_{p}/\xi_{2}\approx 1.215$ GeV is close to the mass of the lightest $ \Delta$ baryons.
Dimensionless parameter $ \xi_{2}$ gives also the    effective threshold values for
the elastic \textit{pp} scattering amplitude, as we will show in a following article.

In conclusion, we do not know why the roots of equation (12), the dimensionless scales $ \xi_{1}, \xi_{2}$, are relevant for hadron physics, but we believe that this observation
may help to find an adequate model for the elastic scattering amplitude.

\section{Discussion}

We have analyzed the behavior of cross-sections of the \textit{pp}
collisions at all energies $ \sqrt{s}\geq 3$ GeV. The main factor determining the behavior of elastic and total cross-sections and their ratio is a monotonous and fast (faster than $ \ln (\sqrt{s} \,)$) growth of the inelastic cross-section $\sigma_{\mathrm{inel} }(s)$ and its large value  compared to $\sigma_{\mathrm{el} }(s)$. Such a rapid growth of $\sigma_{\mathrm{inel} }(s)$ inevitably leads to an increase in $\sigma_{\mathrm{el} }(s)$. Besides, this leads to monotonic growth with energy of the difference $\Delta(s)=\sigma_{\mathrm{inel} }(s)-\sigma_{\mathrm{el} }(s)$. Therefore, the inelastic cross-section grows faster than the elastic one ($\sigma_{\mathrm{inel}}^{'} > \sigma_{\mathrm{el}}^{'}$). But because of  $\Delta(s)>\sigma_{\mathrm{el}}(s)$ the ratio $\sigma_{\mathrm{el}}/\sigma_{\mathrm{tot}} $ and the intensity of inelastic interaction increase at high energies. As a result, the  $\sigma_{\mathrm{tot} }(s), \sigma_{\mathrm{el} }(s)$ and $\sigma_{\mathrm{el}}(s)/\sigma_{\mathrm{tot}}(s) $ have  minima and reach them at energies $\sqrt{s_{\mathrm{tot}}} < \sqrt{s_{\mathrm{el}}} < \sqrt{s_{\mathrm{rat}}}$. So, the  large value and rapid growth of $\sigma_{\mathrm{inel} }(s) $ determine
the overall  picture of the  behavior of the
cross-sections and their ratios (see Fig. 1, Fig. 2 ). This picture has a universal character (see plots for \textit{$\bar{p}$p} , $ \pi p $ and $ K p$ in [21]). Note that all these results confirm the shadow nature of elastic scattering.

Surprisingly, if the minimum  value of the intensity of  inelastic interaction for \textit{pp} collisions  is equal to the value of the ratio $\xi_{\mathrm{1}}\equiv (m_{\pi^{\mathrm{0}}}/ m_{p}) \approx0.143856,$ then the predicted minimum  value of the ratio $\sigma_{\mathrm{el}}/\sigma_{\mathrm{tot}} $ will be very close to the experimental one (see Fig. 2). It may also be that $ \rho(s)\leq \xi_{1}$, see Fig. 3.
In other words, the same number $\xi_{1}$ occurs in dimensionless ratios of many different quantities of hadron physics.
This is similar to the situation with GR in macroscopic phenomena. Therefore, it is not so surprising that $\xi_{1} $ is one of the roots  of the quadratic equation
$(9\,x^{2}+4\,\sqrt{2}\,x-1)=0 $.  The second root of this equation, $-\xi_{2}\approx -0.77239 $, is also related to the hadron mass ratios and gives the   effective threshold values for
the elastic \textit{pp} scattering amplitude, as we will show in a following article. These results are  purely phenomenological observations, we do not know why the scales $ \xi_{1}, \xi_{2}$ are relevant for hadron physics, but we believe that these observations
may help to find an adequate model for the elastic scattering amplitude.

\section{Acknowledgements}
I am grateful to V.A. Petrov  for useful stimulating discussions and the critical remarks and also to A.V. Kisselev, R.A. Ryutin, S.N. Storchak for interesting
discussions.

\end{document}